\documentclass[11pt,a4paper]{article}
\usepackage{jstyle}

\DeclareMathOperator\iso{\mathfrak{iso}}
\DeclareMathOperator\so{\mathfrak{so}}
\DeclareMathOperator\su{\mathfrak{su}}

\DeclareMathOperator\uu{\mathfrak{u}}
\DeclareMathOperator\syp{\mathfrak{sp}}

\def\da{\dot{a}}
\def\db{\dot{b}}
\def\dc{\dot{c}}
\def\dd{\dot{d}}
\def\Lt{\tilde{L}}

\usepackage{tikz-cd}

\title{\centering Spinor-helicity representations of \\
(A)dS$_4$ particles
of any mass}

\author[a]{Thomas Basile}
\emailAdd{thomas.basile@umons.ac.be}
\affiliation[a]{Service de Physique de l'Univers, Champs et Gravitation,  Universit\'e de Mons, 20 place du Parc, 7000 Mons, Belgium}

\author[b]{\quad Euihun Joung}
\emailAdd{euihun.joung@khu.ac.kr}
\affiliation[b]{Department of Physics, College of Science, Kyung Hee University, Seoul 02447, Republic of Korea}

\author[c]{\quad Karapet Mkrtchyan} 
\emailAdd{k.mkrtchyan@imperial.ac.uk}
\affiliation[c]{Theoretical Physics Group, Blackett Laboratory, Imperial College London, SW7 2AZ, UK}

\author[d]{\quad Matin Mojaza\footnote{Matin Mojaza contributed to this work while working at AEI Potsdam until September 2022.}}
\affiliation[d]{Albert-Einstein-Institut, Max-Planck-Institut f\"ur Gravitationsphysik, \mbox{14476 Potsdam,} Germany}

\abstract{
The spinor-helicity representations of massive and (partially-)massless particles in
four dimensional (Anti-) de Sitter spacetime
are studied
within the framework
of the dual pair correspondence.
We show that the dual groups (aka ``little groups'')
of the AdS and dS groups are respectively 
$O(2N)$ and $O^*(2N)$.
For $N=1$, the generator
of the dual algebra  $\mathfrak{so}(2)
\cong \mathfrak{so}^*(2) \cong \mathfrak{u}(1)$ 
corresponds to the helicity operator,
and the spinor-helicity representation
describes massless particles in (A)dS$_4$.
For $N=2$, the dual algebra is
composed of two ideals, $\mathfrak{s}$
and $\mathfrak{m}_\L$.
The former ideal $\mathfrak{s}\cong \mathfrak{so}(3)$ fixes the spin
of the particle,
while the mass is determined by the latter ideal $\mathfrak{m}_\L$,
which is isomorphic to 
$\mathfrak{so}(2,1)$, $\mathfrak{iso}(2)$ or $\mathfrak{so}(3)$
depending on the cosmological constant being positive,
zero or negative. 
In the case 
of a positive cosmological constant, namely dS$_4$,
the spinor-helicity representation
contains all massive particles
corresponding to the principal series 
representations 
and the partially-massless particles corresponding to the discrete series representations
leaving out only
the light massive particles
corresponding to the complementary
series representations.
The zero and negative cosmological
constant cases, which had been
addressed in earlier references,
are also discussed briefly.
Finally, we consider
the multilinear form 
of helicity spinors invariant under (A)dS group, which can be served for
the (A)dS counterpart of the scattering amplitude,
and discuss technical differences
and difficulties of the (A)dS cases
compared to the flat spacetime case.

}

\begin{document}

{\phantom{.}\vspace{-2.5cm}\\\flushright Imperial-TP-KM-2024-01\\ }

\maketitle

\section{Introduction}

The massless spinor-helicity (SH) representation
in flat spacetime (Mink$_4$) has proven very effective 
in expressing and determining scattering amplitudes
(see e.g. \cite{Dixon:1996wi, Elvang:2013cua, Henn:2014yza, Elvang:2015rqa}
for reviews) and their massive counterpart is also prevalent
in recent time (see \cite{Conde:2016vxs, Conde:2016izb}
and \cite{Arkani-Hamed:2017jhn,Maybee:2019jus}, and more).
Moreover, several attempts to generalize it
to (Anti)-de Sitter spaces ((A)dS$_4$) were undertaken
in the literature (see e.g.
\cite{Maldacena:2011nz,David:2019mos, Albrychiewicz:2021ndv} for dS$_4$
and \cite{Nagaraj:2018nxq, Nagaraj:2019zmk, 
Nagaraj:2020sji} for AdS$_4$).
In the latter series of references
\cite{Nagaraj:2018nxq, Nagaraj:2019zmk, Nagaraj:2020sji},
the Mink$_4$ SH representation 
is deformed to (A)dS$_4$ ones with
a term in translation generators proportional to the cosmological constant.
Despite this deformation, the main salient structure
of the scattering amplitude remains the same, while only the momentum conservation delta function is modified to a 
$\Lambda$-dependent function.

In this paper, we generalize 
the (A)dS$_4$ SH representation used in  
\cite{Maldacena:2011nz, Nagaraj:2018nxq, David:2019mos, Nagaraj:2019zmk, Nagaraj:2020sji}
to include massive and partially-massless cases
and carefully analyze their
irreducible representation (irrep) content.
Our analysis is systematic, using the reductive dual pair
correspondence \cite{Howe1989i, Howe1989ii}
(see \cite{Rowe:2011zz, Rowe:2012ym, Basile:2020gqi}
for physics-oriented reviews, and
\cite{Kudla1986, Prasad1993, Adams2007, Sun2020}
for mathematics-oriented ones), the adequate mathematical
framework responsible for most of the technical successes,
yet always behind the curtain in the physicists' treatments
of the subject.

We show that the dual groups 
of the AdS$_4$ and dS$_4$ groups are respectively 
$O(2N)$ and $O^*(2N)$.
For the $N=1$ case, the generator
of the dual algebra  $\mathfrak{so}(2)
\cong \mathfrak{so}^*(2) \cong \mathfrak{u}(1)$ 
corresponds to the standard helicity operator
of the SH formalism,
and the SH representation
describes massless fields\footnote{The SH representations describe single-particle states, but we will use the term ``field'' and ``particle'' interchangeably, as the 
most relevant context is scattering amplitudes in quantum field theory.} in (A)dS$_4$.
For the $N=2$ case, the dual algebra is
composed of two ideals, $\mathfrak{s}$
and $\mathfrak{m}_\Lambda$.
The former ideal $\mathfrak{s} \cong \mathfrak{so}(3)$
fixes the spin of the (A)dS field,
while the mass of the field is determined
by the latter ideal $\mathfrak{m}_\Lambda$,
which is isomorphic to 
$\mathfrak{so}(1,2)$, $\mathfrak{iso}(2)$
or $\mathfrak{so}(3)$ depending on the cosmological
constant being positive, zero, or negative. 
In the case of positive cosmological constant,
namely dS$_4$, the SH representation
contains all massive fields corresponding
to the principal series representations
of $\mathfrak{so}(1,4)$ and the partially-massless
fields corresponding to the discrete series
representations of $\mathfrak{so}(1,4)$.
The only irreps left out are the \emph{light}
massive fields corresponding to the complementary
series representations of $\mathfrak{so}(1,4)$.
We also comment on the Mink$_4$ 
and the AdS$_4$ case, analyzed in earlier 
literature. The Mink$_4$ case was analyzed in details
in the earlier work \cite{Conde:2016izb} of the two
of the authors. See also  more widely known later
 work \cite{Arkani-Hamed:2017jhn}.
The AdS$_4$ case was analyzed in \cite{Basile:2020gqi}
in terms of  creation/annihilation
operators.
We also briefly comment on
the dual pairs
responsible for
the SH representations of (A)dS particles in other dimensions.

Remark that the dual group is also known
as ``little group''.
This terminology
is misleading because
the dual group differs from
the little group of
the induced representation \`a la Wigner:
the actual little group is a subgroup of Lorentz,
while the dual group 
commutes with the Lorentz.
See the Appendix of \cite{Conde:2016izb} for
the explicit comparison
between the little group and dual group in the case of Poincar\'e algebra.

Finally, we consider the multilinear form 
of helicity spinors invariant under (A)dS$_4$ group,
which can be used for
the (A)dS counterpart of the scattering amplitude.
Despite the similarity with the Mink$_4$ case,
we find a few technical differences
and difficulties in the (A)dS$_4$ cases.
We discuss these points and propose potential resolutions.

\section{Spinor-helicity Representations
of (A)dS fields}
The SH representation of massive Mink$_4$ fields 
\cite{Conde:2016izb,Arkani-Hamed:2017jhn}
and that of massless (A)dS$_4$ fields 
\cite{Maldacena:2011nz, Nagaraj:2018nxq, 
Nagaraj:2019zmk,Nagaraj:2020sji} admit 
a common and simple generalization,
\begin{equation} 
    P_{a \dot b}
        = \lambda^I{}_{a}\,\tilde \lambda_{I\,\dot b}
        +\L\,\frac{\partial}{\partial \lambda^{I\,a}}
        \frac{\partial}{\partial \tilde{\lambda}_I{}^{\dot b}}\,,
        \label{sym P}
\end{equation}
\begin{equation}
    L_{ab} = 2\,i\,\lambda^I{}_{(a}\,
            \frac{\partial}{\partial \lambda^{I\,b)}}\,,
    \qquad
    \tilde L_{\dot a\dot b}
        = 2\,i\,\tilde\lambda_{I(\dot a}\,
        \frac{\partial}{\partial \tilde\lambda_I{}^{\dot b)}}\,,
	\label{sym L}
\end{equation}
where $I=1,\ldots, N$\,, and 
the $N=1$ case corresponds to the massless
case and the $\Lambda=0$ limit corresponds to
Mink$_4$ case. Here, $\tilde{\lambda}_{I\dot a}$
is the complex-conjugate of $\lambda^I{}_a$
for real ``momenta''. Round brackets indicate
symmetrization with weight one.
Both of the indices $a,b$ and $\dot a, \dot b$
are raised and lowered by the two-dimensional
Levi--Civita tensor.\footnote{We follow notations
and conventions of \cite{Conde:2016izb} with
\begin{equation}
	\left(\sigma^\mu\right)_{a\dot b}
    =\left(1,\vec{\sigma}\right)_{a\dot b}\,,
    \qquad
	\left(\bar\sigma^\mu\right)^{\dot ab}=
	\epsilon^{\dot a\dot d}\,\epsilon^{bc}\,
	\left(\sigma^\mu\right)_{c\dot d}
    = \left(1,-\vec{\sigma}\right)^{\dot ab}\,,
\end{equation}
where $\sigma^i$, $i=1,2,3$, are the usual
Pauli matrices, which verify
$(\sigma^\mu)_{a \dot a}\,(\sigma_\mu)_{b \dot b}
= -2\,\epsilon_{a \dot a}\,\epsilon_{b \dot b}$,
and
\begin{equation}
	(\sigma^{\mu\nu})_a{}^b=
	\frac14\,(\sigma^\mu\,\bar{\sigma}^\nu
    -\sigma^\nu\,\bar{\sigma}^\mu)_a{}^b\,,
    \qquad
	(\bar\sigma^{\mu\nu})^{\dot a}{}_{\dot b}
    =-\frac14\,(\bar\sigma^\mu\,\sigma^\nu
    -\bar\sigma^\nu\,\sigma^\mu)^{\dot a}{}_{\dot b}\,.
\end{equation}
Indices are raised and lowered via
\begin{equation}
	\psi_a = \epsilon_{ab}\,\psi^{b}\,\,,
	\quad \psi^a=\epsilon^{ab}\,\psi_{b}\,,
	\quad\quad \epsilon^{ac}\,\epsilon_{cb}=\d^a_b\,,
\end{equation}
and similarly for dotted indices.}
We shall denote this (A)dS$_4$ isometry algebra
as $\mathfrak{sym}_\Lambda$. It is straightforward
to check that the commutators of the above operators
satisfy the Lie brackets of the (A)dS$_4$ algebra
with cosmological constant $\Lambda$: the generators
$L_{ab}$ and $\tilde L_{\dot a \dot b}=L_{ab}{}^\dagger$
form standard Lorentz subalgebra
$\mathfrak{so}(1,3) \cong \mathfrak{sl}(2,\mathbb C)$
with $[L_{ab},\tilde L_{\dc\dd}]=0$ and 
\begin{align}
    [L_{ab}, L_{cd}] = -i\,(\epsilon_{ac}\,L_{bd}
    + \epsilon_{bc}\,L_{ad} +\epsilon_{ad}\,L_{bc}
    + \epsilon_{bd}\,L_{ac})\,.
\end{align}
The translation generators $P_{a\dot b}$ carry a vector representation of $\mathfrak{so}(1,3)$,
that is a bifundamental representation
of $\mathfrak{sl}(2,\mathbb C)$,
\begin{equation}
    [L_{ab}, P_{c \dot d}] = i\,(\epsilon_{ca}\,P_{b \dot d}
    + \epsilon_{cb}\,P_{a \dot d})\,,
    \qquad 
    [\tilde L_{\dot a \dot b}, P_{c \dot d}]
    = i\,(\epsilon_{\dot d \dot a}\,P_{c \dot b}
    + \epsilon_{\dot d \dot b}\,P_{c \dot a})\,.
\end{equation}
With the cosmological constant $\Lambda$,
the translation generators no longer commute but
satisfy
\begin{equation}\label{Poincare}
    [P_{a \dot b}, P_{c \dot d}] = i\,\Lambda\,
    (\epsilon_{ac}\,\tilde L_{\dot b \dot d}
        + \epsilon_{\dot b \dot d}\,L_{ac})\,.
\end{equation}
Hence, we find
$\mathfrak{sym}_\Lambda \simeq \so(1,4)$ for $\Lambda>0$
and 
$\mathfrak{sym}_\Lambda \simeq \so(2,3)$ for $\Lambda<0$\,.\footnote{
Note here that $\L$ is
related to the actual cosmological constant $\L_{cc}$ by $\L_{cc}=3\,\L$.}

The (A)dS$_4$ algebra $\mathfrak{sym}_\Lambda$
is a subalgebra of $\syp(8N,\mathbb R)$
generated by all bilinears in 
$\lambda^I_a$, $\frac{\partial}{\partial\lambda^I_a}$
and their complex conjugates. The dual algebra,
denoted by $\mathfrak{dual}^{(N)}_\Lambda$,
is the stabiliser of $\mathfrak{sym}_\Lambda$ 
within $\syp(8N,\mathbb R)$, and is generated by
\begin{subequations}
\label{dual alg}
\begin{equation}
    K^I{}_J = \lambda^I{}_a\,\frac{\partial}{\partial \lambda^J{}_a}
    - \tilde\lambda_{J \dot a}\,
            \frac{\partial}{\partial\tilde\lambda_{I\,\dot a}}\,,
    \label{dual K}
\end{equation}
\begin{equation}
	M^{IJ} = \lambda^I{}_{a}\,\lambda^{Ja}
    -\Lambda\,\frac{\partial}{\partial\tilde{\lambda}_I{}^{\dot a}}
    \frac{\partial}{\partial \tilde{\lambda}_{J\,\dot a}}\,,
    \qquad 
	\tilde M_{IJ} =	\tilde\lambda_{I\dot a}\,
                        \tilde\lambda_J{}^{\dot a}
    -\Lambda\,\frac{\partial}{\partial \lambda^{I\,a}}
    \frac{\partial}{\partial \lambda^J{}_a}\,.
	\label{dual M}
\end{equation}
\end{subequations}
The SH representation of $\mathfrak{sym}_\L$ is reducible
and its decomposition into irreps
can be carried out on
the side of $\mathfrak{dual}_\L^{(N)}$.
In the following,
 we shall identify the 
 dual algebra
$\mathfrak{dual}^{(N)}_\Lambda$
and explain the intimate relation between 
$\mathfrak{sym}_\Lambda$ and 
$\mathfrak{dual}^{(N)}_\Lambda$,
first through a preliminary analysis 
on the eigenvalues of Casimir operators,
then using the more solid
and powerful method of the dual pair
correspondence.

\section{Preliminary analysis}
In this section, we identify the dual algebra 
$\mathfrak{dual}^{(N)}_\Lambda$ for $N=1, 2$,
and establish its relation to  $\mathfrak{sym}_\Lambda$
at the level of Casimir operators.
By comparing the eigenvalues of the Casimir operators 
of $\mathfrak{sym}_\Lambda$ and $\mathfrak{dual}^{(N)}_\Lambda$, 
we provide a preliminary assessment
of the correspondence between
the irreps of 
$\mathfrak{sym}_\Lambda$
and $\mathfrak{dual}^{(N)}_\Lambda$\,.

\subsection{$N=1$}

In the $N=1$ case, considered in 
\cite{Nagaraj:2018nxq, Nagaraj:2019zmk, Nagaraj:2020sji}, the dual algebra
$\mathfrak{dual}^{(1)}_\Lambda$ is simply isomorphic to
$\mathfrak{u}(1)$ generated by
\be 
    K=\lambda_a\,\frac{\partial}{\partial \lambda_a}-
    \tilde\lambda_{\dot a}\,\frac{\partial}{\partial \tilde\lambda_{\dot a}}\,,
\ee 
which is nothing but the standard helicity operator.
The $K=s$ state describes massless 
helicity $s$ representations
in Mink$_4$, AdS$_4$ and dS$_4$.
This universal description
is due to the conformal symmetry
they enjoy: the SH representations 
of $\mathfrak{sym}_\L$ can be lifted
to a single irreducible representation,
typically referred to as `singleton',
of the four-dimensional conformal group $\so(2,4)$
\cite{Angelopoulos:1980wg, Angelopoulos:1997ij, 
Laoues:1998ik, Angelopoulos:1999bz}
(see also \cite{Sezgin:2001yf, Sezgin:2001zs,
Gunaydin:1998sw}
for the oscillator realization,
where sometimes
the representation is referred to as 
`doubleton' for a historical reason).
This special property of singleton
can be easily understood in terms of the dual pair correspondence, as it was shown
in \cite{Basile:2020gqi}. 
We shall come back to this point in Section \ref{sec:singletons}.

\subsection{$N=2$}
The $N=2$ case will turn out to be sufficient
to describe all massive spin representations
in four dimensions. The generators $M=M^{12}$
and $\tilde M=\tilde M_{12}$ commute with the subalgebra
$\so(3) \simeq \su(2) \subset \uu(2)$ generated by 
$\cK^I{}_J = K^I{}_J - \frac12\,\delta^I_J\,K^K{}_K$
while the $\uu(1)$ {part $K=K^I{}_I$} satisfies
\begin{equation}
    [M, \tilde{M}] = -\Lambda\,K,
    \qquad
	[K,M] = 2\,M\,,
    \qquad
    [K,\tilde M] = -2\,\tilde M\,.
\end{equation}
Taking into account that
$M^\dagger=\tilde{M}$ and $K^\dagger=K$,\footnote{Note 
that the Hermitian conjugation $\dagger$ is defined with respect
to the $L^2(\mathbb C^{2N})$ norm,
and hence
$\lambda^I{}_a{}^\dagger=
(\lambda^I{}_a)^*=\tilde\lambda_{I\,\da}$
and $(\partial/\partial\lambda^I{}_a)^\dagger=-\partial/\partial\tilde\lambda_{I\,\da}$.}
it is easy to show that the Hermitian generators
$\frac12\,K$, $\frac12\,(M+\tilde{M})$
and $\frac{i}2\,(M-\tilde M)$ 
form 
$\so(2,1)$ for $\Lambda>0$, $\so(3)$ for $\Lambda<0$
and $\iso(2)$ for $\Lambda=0$. The last case
corresponds to the massive Mink$_4$ SH formulation \cite{Conde:2016izb,Arkani-Hamed:2017jhn}.
To summarize, we find that for $N=2$,
the dual algebra is the direct sum,
\begin{equation}
    \mathfrak{dual}^{(2)}_\Lambda \simeq
    \mathfrak{s} \oplus \mathfrak{m}_\Lambda\,,
\end{equation}
where the two ideals $\mathfrak{s}$
and $\mathfrak{m}_\Lambda$ are 
\begin{equation}
    \mathfrak{s}=\mathfrak{so}(3)\,,
    \quad 
    \mathfrak{m}_\L = \left\{
    \begin{aligned}
        & \mathfrak{so}(2,1)
        \quad & [\L>0]\\
        & \mathfrak{so}(3)
        \quad & [\L<0]\\
        & \mathfrak{iso}(2)
        \quad & [\L=0]
    \end{aligned}
    \right..
\end{equation}
Below, we will show that the common
ideal $\mathfrak{s}$ for any $\Lambda$
is responsible for the spin label
of the $\mathfrak{sym}_\L$ irreps,
whereas  the other subalgebra
$\mathfrak{m}_\Lambda$ determines the mass.
In order to see this identification,
let us first exploit the relations between Casimir 
operators of $\mathfrak{sym}_\Lambda$
and $\mathfrak{dual}^{(2)}_\Lambda$.

Since $\mathfrak{sym}_\Lambda$ is a rank two Lie algebra
$\mathfrak{so}(1,4)$ or $\mathfrak{so}(2,3)$ for $\L\neq 0$,
there are two independent Casimir operators: the quadratic
and quartic ones, whose  expressions in vector notation
read 
\begin{subequations}
\begin{eqnarray}
    && C_2(\mathfrak{sym}_\Lambda)
    = -\frac12\, J^{A_1}{}_{A_2}\,J^{A_2}{}_{A_1}\,, \\
    &&C_4(\mathfrak{sym}_\Lambda)
    = \frac12\, W_A\, W^A\,,\qquad W_A=\frac12\,\epsilon_{ABCDE} \,J^{BC}\,J^{DE}\,,
\end{eqnarray}
\end{subequations}
where the capital indices take the values
$A,B,\dots=0,1,\dots,4$ and $J_{AB}=-J_{BA}$
are the generators of $\mathfrak{sym}_\Lambda$.
Splitting $J_{AB}$ into Lorentz and translation generators
as $J_{4\mu}=P_{\mu}/\sqrt{|\Lambda|}$
and $J_{\mu\nu} = L_{\mu\nu}$, 
the two Casimirs are 
\begin{subequations}
\begin{eqnarray}
    C_2(\mathfrak{sym}_\Lambda)
    \eq -\frac1{2\,\Lambda}\,P^2
    + \frac14\,(L^2 + \tilde{L}^2)\,,\\
    C_4(\mathfrak{sym}_\Lambda)
    \eq \frac1{4\,\Lambda}\,P^{a \dot a}\,
    P^{b \dot b}\,L_{ab}\,\tilde L_{\dot a \dot b}
    + \frac1{16\,\Lambda}\,P^2\,(L^2 + \tilde L^2)\nn 
    && -\,\frac14\,(L^2 + \tilde L^2)
    -\frac1{64}\,(L^2 - \tilde L^2)^2 \,.
\end{eqnarray}
\end{subequations}
Here, $P^2=P_{a\dot b}\,P^{a\dot b}=-2\,P_{\mu}P^{\mu}$, $L^2=L_{ab}L^{ab}$ and $L_{\mu\nu}L^{\mu\nu}=\frac12(L^2+\Lt^2)$, where we use the mostly-plus signature for $\eta_{\mu\nu}$. Note that
$\Lambda\,C_2(\mathfrak{sym}_\Lambda)$
and $\Lambda\,C_4(\mathfrak{sym}_\Lambda)$
reproduce the familiar quadratic Casimir
and the Pauli--Luba\'nski vector squared 
in the $\Lambda \to 0$ limit.

On the other hand, the dual algebra
is composed of two rank-one ideals,
so we have one Casimir operator for each:
\begin{eqnarray}
    C_2(\mathfrak{s}) \eq \frac12 \cK^I_J\,\cK^J_I\,,\\ 
    C_2(\mathfrak{m}_\Lambda) \eq -\frac1{2\Lambda}\,
    \{M, \tilde M\} + \frac14\,K^2\,.
\end{eqnarray}
The SH representation
of $\mathfrak{sym}_\Lambda$ \eqref{sym P}
and \eqref{sym L}
and that of $\mathfrak{dual}^{(2)}_\Lambda$ 
\eqref{dual K} and \eqref{dual M}
relate these Casimir operators as
\begin{subequations}
\begin{eqnarray}
    C_2(\mathfrak{sym}_\Lambda)
   \eq  C_2(\mathfrak{m}_\Lambda)
   + C_2(\mathfrak{s}) - 2\,, \\
    C_4(\mathfrak{sym}_\Lambda)
    \eq -\,C_2(\mathfrak{m}_\Lambda)\,C_2(\mathfrak{s})\,.
\end{eqnarray}
\end{subequations}
From the above relations,
we can read off the Casimir 
eigenvalues of the unitary irreps
of $\mathfrak{sym}_\Lambda$ by fixing an irrep
of $\mathfrak{dual}^{{(2)}}_\Lambda
\simeq \mathfrak{s} \oplus \mathfrak{m}_\Lambda$.
For the ideal $\mathfrak{s} \simeq \mathfrak{so}(3)$,
the $(2s+1)$-dimensional irreps with
\begin{equation}
    C_2(\mathfrak{s}) = s(s+1)\,,
\end{equation}
account for all unitary irreps.
About the ideal $\mathfrak{m}_\Lambda$,
the quadratic Casimir operator
can be parameterized as
\begin{equation}
     C_2(\mathfrak{m}_\Lambda) = \mu(\mu+1)\,,
\end{equation}
which is invariant under 
\be
    \mu \to -1-\mu,
    \label{shadow}
\ee
and we have the following options:
\begin{itemize}
\item For $\Lambda>0$, apart from the trivial irrep with $\mu(\mu+1)=0$,
we have three series of unitary irreps for $\mathfrak{so}(2,1)$\,:
    \begin{itemize}
        \item The principal series irreps $\cC_\mu^{\pm}$
        with complex $\mu$ satisfying
       \begin{equation}
            \mu(\mu+1) < -\frac14\,,
            \label{principal}
        \end{equation}
        which is spanned by eigenstates
        of $K$ with even/odd integer eigenvalues,
        related to the label $+/-$ respectively.
        We can parametrize irreps in this series
        via $\mu=-\frac12+i\,\rho$
        with $\rho\in\mathbb R$.
        In this case, the map \eqref{shadow},
        $\rho\to -\rho$, is an isomorphism,
        and hence we may restrict
        to the case $\rho>0$.

        \item The complementary series irrep $\cC_\mu$
        with $-1< \mu < 0$ satisfying
        \begin{equation}
             -\frac14 \leq \mu(\mu+1) < 0\,,
            \label{complementary}
        \end{equation}
        spanned by all even $K$-eigenstates.
        The map \eqref{shadow}
        is again an isomorphism.
        
        \item The positive/negative discrete series
        irrep $\cD^\pm_{2\mu+2}$  with
        \begin{equation}
            \mu = -\tfrac12, 0, 
            \tfrac12, 1, \tfrac32,  \ldots\,,
            \label{discrete}
        \end{equation}
         spanned by the $K$-eigenstates with eigenvalues
         $\pm 2(\mu+1), \pm 2(\mu+2)$, etc.
         These are lowest/highest weight irreps.
         
    \end{itemize}
\item For $\Lambda \to 0$\,, the ``bosonic/fermionic''
irrep of $\iso(2)$ with $|\mu|\to\infty$
while keeping finite
\begin{equation}
    m ={\sqrt{-\Lambda\,\mu^2}}\,,
\end{equation}
which is spanned by $K$-eigenstates
with even/odd eigenvalues. These irreps
can be thought of as the counterpart of the massive
scalar and spinor representations of the Poincar\'e
group (depending on the parity
of the $K$-eigenstates).

The trivial representation, with $m=0$,
and which can be thought of as the counterpart
of the zero-momentum irrep of the Poincar\'e group.

\item For $\Lambda<0$, the $(2\mu+1)$-dimensional irrep
of $\so(3)$ with 
\begin{equation}
    \mu = 0, \tfrac12, 1, \tfrac32, \ldots\,,
    \label{finite}
\end{equation}
with basis composed of $K$-eigenstates
with eigenvalues $-2\mu,-2\mu+2,\dots,+2\mu$.
\end{itemize}

These irreps of $\mathfrak{dual}^{{(2)}}_\Lambda$
are in one-to-one correspondence with the irreps
of $\mathfrak{sym}_\Lambda$ with
\begin{subequations}
\begin{align}
    C_2(\mathfrak{sym}_\Lambda)
    & = \mu(\mu+1)+s(s+1)-2\,,\\
    C_4(\mathfrak{sym}_\Lambda)
    & = -\mu(\mu+1)\,s(s+1)\,,
    \label{Casimir ev}
\end{align}
\end{subequations}
and we can compare these values with those
of known irreps of $\mathfrak{sym}_\Lambda$.

\subsubsection*{Mink$_4$}
To begin with, let us consider the Poincar\'e 
case with $\Lambda=0$ which has been treated
in \cite{Conde:2016vxs, Arkani-Hamed:2017jhn}.
The quadratic Casimir,
\be
    \lim_{\L\to0} \L\,C_2(\mathfrak{m}_{\L})=
    -M\,\tilde M\,,
\ee
of the dual algebra $\mathfrak{m}_0$
determines the mass:
\be
    M\,\tilde M=-P_\mu\,P^\mu=m^2\,,
\ee
while the `spin $s$' representation of 
the dual algebra $\mathfrak{s}$ corresponds to the spin,
thus defining a Poincar\'e representation of mass $m$
and spin~$s$\,.
In fact, in all cases of $\mathfrak{sym}_\Lambda$,
the irrep label $s$ of the dual algebra $\mathfrak{s}$ 
simply corresponds to the spin of the four-dimensional
field.

\subsubsection*{dS$_4$}
The unitary irreps of dS$_4$ Lie algebra,
namely $\mathfrak{so}(1,4)$,
were first classified in \cite{Dixmier1961}
where the eigenvalues of the Casimir operators
are also given: see Appendix \ref{app:dixmier}
for a summary,
and \cite{Francia:2008hd} for the physical
interpretations of these irreps.  
More recent treatments of dS representations
can be found e.g. in
\cite{Joung:2006gj, Joung:2007je, Basile:2016aen, 
Sun:2021rrs, Sun:2021thf, Enayati:2022hed}.

Comparing the result \eqref{Casimir ev}
with the Casimir eigenvalues identified
in  \cite{Dixmier1961},
we find that the irrep label $\mu$ 
of the dual algebra $\mathfrak{m}_\L$ parameterizes
the mass squared as\footnote{Here, we define
the mass $m^2$ of a field $\varphi$ of spin $s$
in (A)dS$_{d+1}$ via the wave equation
$$\left(\nabla^2 + \frac{2\,\Lambda_{cc}}{d(d-1)}\,
\big[(s-2)(s+d-2)-s\big] - m^2\right)\varphi = 0\,.$$
Parameterizing the eigenvalue of the quadratic
Casimir operator of the irrep associated with
$\varphi$ as 
$$C_2=\Delta(\Delta-d)+s(s+d-2)\,,$$ we can write 
the mass squared as
$$m^2 = \frac{2\,\Lambda_{cc}}{d(d-1)}\,
(\Delta+s-2)(s+d-2-\Delta)\,,$$
which reproduces the formula \eqref{mass squared}
upon using $\mu=\Delta-2$ (or $\mu=-\Delta+1$)
for $d=3$. (See for instance \cite{Garidi:2003ys} 
for an extended discussion of the dS$_4$ case, 
and \cite{Gazeau:2006uy, Gazeau:2008zz}
including also the AdS$_4$ case.)}
\begin{equation}
    m^2 = \L\,[-\mu(\mu+1)+s(s-1)]\,.
    \label{mass squared}
\end{equation}
Depending on the spin $s$, different ranges
of mass are allowed for the unitarity
of the $\mathfrak{sym}_\Lambda$ irreps:
\begin{itemize}
    \item For the scalar case with $s=0$,
    the allowed $\mu$ are
    \begin{itemize}
    \item 
    The complex values of $\mu$ with
    \eqref{principal}
    corresponding to
    the principal 
    series representations of $\mathfrak{so}(1,4)$,
    with the isomorphism \eqref{shadow}.
    \item 
    The real values of $-2< \mu<1$ with
    \be
        -\frac14\le \mu(\mu+1)<2\,,
    \ee
    corresponding to
    the complementary
    series representations of $\mathfrak{so}(1,4)$,
    with the isomorphism \eqref{shadow}.
    The $\mu=0$ case (or equivalently,
    the $\mu=-1$ case) corresponds
    to the conformally coupled scalar.
    \item 
    The positive integer values of $\mu$
    corresponding to
    the discrete series representations
    of $\mathfrak{so}(1,4)$. 
    The $\mu=1$ case corresponds 
    to the minimally coupled massless scalar,
    whereas $\mu=2,3,\ldots$
    correspond to tachyonic scalars.
     \end{itemize}
    The unitarity of these $\mathfrak{sym}_\Lambda$
    irreps includes not only
    all the $\mathfrak{m}_\Lambda$ unitary regions
    \eqref{principal}, \eqref{complementary}
    and the integer part of \eqref{discrete},
    but also the complementary series
    region $0<\mu(\mu+1)<2$
     not allowed for the unitarity
    of $\mathfrak{m}_\Lambda$\,.

    \item For integral spins $s=1,2,\ldots$,
    the allowed $\mu$
    are 
    \begin{itemize}
    \item 
    The complex values
    with \eqref{principal}
    corresponding to
    the principal series representations
    of $\mathfrak{so}(1,4)$,
    with the isomorphism \eqref{shadow}.
    \item 
    The real values of
     $-1 <\mu<0$ with
    \eqref{complementary}
    corresponding to
    the complementary series 
    representations
    of $\mathfrak{so}(1,4)$,
    with the isomorphism \eqref{shadow}.
    \item 
    The integer values $\mu=0,1,\ldots, s-1$.
    These integer values correspond
    to the partially-massless fields of
    depth $s-\mu$, where the depth 1 corresponds to the massless field.
\end{itemize}
    The unitarity 
    of these $\mathfrak{sym}_\L$ irreps  includes
    the $\mathfrak{m}_\L$  unitary regions \eqref{principal} and \eqref{complementary},
    but restrict \eqref{discrete}:
    any integers greater than $s-1$
    are excluded together with
    the half-integer values.

    \item For half-integral spins $s=\frac12,\frac32,\ldots$,
    the allowed $\mu$
    are 
    \begin{itemize}
        \item 
   The complex values of $\mu$
    with \eqref{principal}
    corresponding to the principal
    series representation of $\mathfrak{so}(1,4)$,
    with the isomorphism \eqref{shadow}.
    \item 
    The half integer values $\mu= -\frac12,\frac12,\ldots, s-1$
    corresponding 
    to the discrete series representations 
    of $\mathfrak{so}(1,4)$.
    The positive half-integer values correspond
    to the partially-massless fields of
    depth $s-\mu$.\footnote{The partially-massless 
    fermion irreps are unitary only in dS$_4$ 
    \cite{Letsios:2023qzq}.}
    Note that $\mu=-\frac12$ corresponds
    to the end point
    of the continuous spectrum of massive fields,
    which we may refer
    to as \emph{the lightest massive fermions}.
    For $s=\frac12$, it simply corresponds to the massless spinor.
    
     \end{itemize}
   The unitarity 
    of these $\mathfrak{sym}_\L$ irreps  includes
    the $\mathfrak{m}_\L$  
    principal series \eqref{principal} but
    entirely excludes the complementary series \eqref{complementary},
    and restrict 
    the discrete series \eqref{discrete}:
    any half-integers greater than $s-1$
    are excluded together with
    the integer values. 
\end{itemize}

\subsection*{AdS$_4$}
In the AdS$_4$ case with $\Lambda<0$,
the irrep label $\mu$ 
of the dual algebra $\mathfrak{m}_\Lambda$ parameterizes
the mass squared again as \eqref{mass squared}.
The allowed $\mu$ for the unitarity of 
the lowest-energy irreps of $\mathfrak{sym}_\Lambda$ are
 $\mu=s-1,s,s+1,\ldots$
 for spin $s=0,\frac12\,1,\ldots$.
The $\mu=s-1$ case corresponds 
to the massless spin $s$ field,
and higher $\mu$ cases correspond
to massive fields.
The reason that we have a discrete mass spectrum
is due to the
fact that $\mu$ is an eigenvalue of the generator
of the \emph{compact} subgroup $SO(2)$ associated
with rotations in the plane of temporal directions,
and hence is quantized.
These representations can be interpreted
as the irreps of 3d conformal group:
$\Delta=\mu+2$ and $s$ correspond to the conformal weight
and spin of the conformal primaries, respectively.
In the scalar case,
the $\mu=-1$ and  $\mu=0$ cases
mapped by \eqref{shadow}
are distinct irreps
and correspond to 
different modes of the conformal scalar in AdS$_4$. 
Note that, 
moving to a 
covering group of 
$SO(1,4)$,
the point $\m=-\frac32$ can be included for $s=0$, and it corresponds
to the conformal scalar in 3d.

The unitarity of the lowest energy irreps
of $\mathfrak{sym}_\Lambda \cong \mathfrak{so}(2,3)$ 
excludes the lower $\mu$ values with $\mu<s-1$
from \eqref{finite}, corresponding to
partially-massless fields, together with
all integer/half-integer values of $\mu$
for half-integral/integral spin.

Let us note that there are a few other types of $\mathfrak{sym}_\L$ irreps
with unbounded energy.
These irreps would cover different ranges
of $C_2(\mathfrak{sym}_\L)$
and $C_4(\mathfrak{sym}_\L)$\,.

\section{Dual pair correspondence}

In the previous section we have identified the correspondences
between the irreps of $\mathfrak{sym}_\Lambda$
and those of $\mathfrak{dual}^{{(2)}}_\Lambda$
through the Casimir eigenvalues.
We have  observed 
that the  region of $\mu$ allowed
by the $\mathfrak{sym}_\Lambda$ unitarity 
does not match the region allowed
by the $\mathfrak{m}_\Lambda$ unitarity.
This mismatch does not lead to a contradiction,
because the SH representations
cover only a part of unitary irreps
of $\mathfrak{sym}_\Lambda
\oplus \mathfrak{dual}^{{(2)}}_\Lambda$.
In other words, the SH Fock space
contains only a part of unitary irreps
of $\mathfrak{sym}_\Lambda
\oplus \mathfrak{dual}^{{(2)}}_\Lambda$\,.
In order to identify the actual content
of the unitary irreps that the Fock space
contains, we need a more rigorous analysis  using
the dual pair correspondence.

For general $N$, the dual algebras \eqref{dual alg}
are $\mathfrak{dual}^{{(N)}}_{\Lambda>0}
\simeq \so^*(2N)$
and $\mathfrak{dual}^{{(N)}}_{\Lambda<0}
\simeq \so(2N)$, respectively.
The interplay between the isometry 
and the dual algebras can be understood within
the general framework of the dual pair correspondence,
aka Howe duality, which amounts to  the following:
when a $Sp(2\cN,\mathbb R)$ group 
contains a pair of reductive subgroups $(G,\tilde G)$ 
which are mutual stabilisers,
there exists a one-to-one correspondence
between the irreps of $G$ and $\tilde G$ appearing
in the decomposition of the oscillator 
(or metaplectic) representation of $Sp(2\cN,\mathbb R)$
(see e.g. \cite{Basile:2020gqi} for more details).
In our context, the oscillator representation 
is simply the representation realized by 
the helicity spinors,
or simply SH representation.
Hence, the (A)dS$_4$ groups
$\mathfrak{Sym}_{\Lambda>0}=Sp(1,1)$
and $\mathfrak{Sym}_{\Lambda<0}=Sp(4,\mathbb R)$
and their respective dual groups
$\mathfrak{Dual}^{{(N)}}_{\Lambda>0}=O^*(2N)$ 
and $\mathfrak{Dual}^{(N)}_{\Lambda<0}=O(2N)$ 
realized by helicity spinors as \eqref{Poincare}
and \eqref{dual alg} form reductive dual pairs
in $Sp(8N,\mathbb R)$, the group generated by
all quadratic operators in helicity spinors
and their derivatives.
Note that $Sp(1,1)$ and $Sp(4,\mathbb R)$ are isomorphic
to the double covers of $SO^{\uparrow}(1,4)$
and $SO^{\uparrow}(2,3)$, respectively.
In fact, the flat space case with $\Lambda=0$
can be viewed as the In\"on\"u--Wigner contraction
of the reductive dual pair $\big(Sp(1,1),O^*(2N)\big)$
or $\big(Sp(4,\mathbb R), O(2N)\big)$.

Let us remark once again that the dual group ought not
to be confused with the standard little group
of the induced representation \`a la Wigner:
the former commutes with the isometry whereas the latter
is a part of the isometry by definition.
In the $\Lambda=0$ case,  the $SU(2)$ subgroup
of the dual group and the little group are explicitly
shown to be distinguished (see the appendix
of \cite{Conde:2016izb}) as they represent respectively
left and right actions on $SU(2)$ which parameterizes
a momentum eigenstate.

The dual pair correspondence assures that 
the irreps of the (A)dS$_4$ group,
that is $Sp(1,1)$ and $Sp(4,\mathbb R)$,
realized by helicity spinors
are in one-to-one correspondence
with the irreps of the dual group
$O^*(2N)$ or $O(2N)$.
In other words, by singling out
an irrep of the dual group,
the reducible SH
representation of the (A)dS$_4$
group \eqref{Poincare}
is restricted to an irrep.
Then, the remaining task is to establish the dictionary between such
irreps of the (A)dS$_4$ group
and its dual group $O^*(2N)$ (or $O(2N)$).
For that, we once again focus
on the cases of $N=1$ and $2$.

\subsection{dS$_4$}
Let us consider first the case with $\Lambda>0$.
Our aim is to obtain a dictionary
between the irreps of $Sp(1,1)$ and $O^*(2N)$ appearing
in the decomposition of the SH representation.

For $N=1$, the dual pair correspondence
between $Sp(1,1)$ and $O^*(2)$ has been explicitly
established in \cite{Basile:2020gqi}.
Here, we just quote the result.
Since $O^*(2)$ is isomorphic to $U(1)$,
it has only one-dimensional irreps,
each labelled by an integer. This integer corresponds
to twice the helicity of a $Sp(1,1)$ massless representation.
The analysis is based on the decomposition
of the $Sp(1,1)$ irrep into its maximal subgroup
$Sp(1) \times Sp(1)$, and the SH representation
restricted by the $O^*(2)$ irrep condition
is shown to have the structure
of the massless spin $s$ irrep of $Sp(1,1)$
demonstrated e.g. in \cite{Dixmier1961}.

For the $N=2$ case, we need
to begin with identifying irreps of the dual group $O^*(4)$.
Thanks to the isomorphism
$O^*(4) \cong [SU(2) \times SL(2,\mathbb R)]/\mathbb Z_2$
(here, $SU(2)$ and $SL(2,\mathbb R)$
are simply the Lie groups associated with 
$\mathfrak{s}=\mathfrak{so}(3)$ and 
$\mathfrak{m}_{\Lambda>0}=\mathfrak{so}(2,1)$),
we know everything about
the unitary irreps of $O^*(4)$:
the irreps of $SU(2)$ are all given by
$(2s+1)$-dimensional representation,
which will be denoted by $[2s]$ henceforth,
while $SL(2,\mathbb R)$
has three classes of unitary irreps,
namely $\cC^\pm_{\mu=-\frac12+i\,\rho}$
\eqref{principal}, $\cC_\mu$ \eqref{complementary}
and $\cD^\pm_{2\mu+2}$ \eqref{discrete}.
We will denote these $O^*(4)$ irreps
as $\tilde\pi_{s,\mu}$.

In the previous section,
we have seen that not all $O^*(4)$ irreps
correspond to irreps
of $Sp(1,1)$ based on the match of Casimir operators.
We shall see below how they are restricted. 
For that, we first consider the dual pair
$\big(Sp(1),O^*(4)\big) \subset Sp(8,\mathbb R)$,
whose representations are  explicitly identified in
\cite[Sec. 5.4]{Basile:2020gqi}:
Since $Sp(1)\cong SU(2)$ the $Sp(1)$ irreps
are again given by $[m]$ with non-negative integer $m$,
and they correspond to the $O^*(4)$ irreps
$[m] \otimes \cD^\pm_{m+2}$\,.
Note that only discrete series representations appear
in the $SL(2,\mathbb R)$ side,
with the highest/lowest weight $m+2$ tied
with the dimension $m+1$ of the $SU(2)$ irrep
(which is  a consequence of the fact
that the Howe dual is a compact group,
namely $Sp(1)$). Whether the irrep $\cD^\pm_{m+2}$
is a highest/lowest weight one is conventional at this stage,
and only one sign is chosen
depending on the convention
of $SL(2,\mathbb R)$.

Now we move on to the  dS$_4$ group $Sp(1,1)$
and consider its maximal compact subgroup,
which is $Sp(1) \times Sp(1)$.
This subgroup 
forms its own dual pair in the same SH space (that is, in $Sp(16, \mathbb R)$) with
$O^*(4) \times O^*(4)$.
The latter contains
the original dual group $O^*(4)$ as the
diagonal subgroup.
The situation is
conveniently depicted by the
``seesaw'' diagram,
\begin{equation}
\parbox{190pt}{
\begin{tikzpicture}
\draw [<->] (0,0) -- (1,0.8);
\draw [<->] (0,0.8) -- (1,0);
\node at (-1.5,1.2) {$Sp(1,1)$};
\node at (-1.5,0.4) {$\cup$};
\node  at (-1.5,-0.4) {$Sp(1)\times Sp(1)$};
\node at (2.5,-0.4) {$O^*(4)$};
\node at (2.5,0.4) {$\cup$};
\node at (2.5,1.2) {$O^*(4)\times O^*(4)$};
\end{tikzpicture}}
\label{seesaw}
\end{equation}
where the arrows indicate the respective dual pairs.
Any irrep of $Sp(1,1)$, say $\pi_\sigma$
with some label $\sigma$, can be decomposed into
irreps of $Sp(1)\times Sp(1)$ as 
\begin{equation}
    \pi_\sigma = \bigoplus_{m,n} N_\sigma^{m,n}\,
    [m] \otimes [n]\,,
    \label{dec}
\end{equation}
where $N^{m,n}_\sigma$ are the multiplicities 
of $[m]\otimes [n]$, and each of $[m]\otimes [n]$
correspond 
to the $O^*(4)\times O^*(4)$ irrep,
\begin{equation} 
    \Big([m] \otimes \cD^-_{m+2}\Big)
    \otimes 
    \Big([n]\otimes \cD^+_{n+2}\Big)\,.
    \label{OO}
\end{equation}
Here, we used the correspondence between the irreps
of $Sp(1)$ and $O^*(4)$ that we introduced earlier.
Note that the first $SL(2,\mathbb R)$ irrep
is a lowest-weight irrep, while the second
is a heighest-weight irrep. This is because
the $Sp(1) \times Sp(1)$ is embedded
in  the opposite signature parts of $Sp(1,1)$.
The irrep \eqref{OO} of $O^*(4)\times O^*(4)$
can be decomposed as well
into the diagonal subgroup $O^*(4)$\,:
\begin{equation}
    \Big([m] \otimes \cD^-_{m+2}\Big)
    \otimes 
    \Big([n]\otimes \cD^+_{n+2}\Big)
    = \bigoplus_{s,\mu} \tilde N^{s,\mu}_{m,n}\,
    \tilde\pi_{s,\mu}\,,
    \label{dual dec}
\end{equation}
where $\tilde N^{s,\mu}_{m,n}$
are the multiplicities of
the $O^*(4)$ irrep $\tilde\pi_{s,\mu}$
that we have introduced before.
The crucial point assured by
the seesaw duality 
(see \cite{Kudla1986, Prasad1993, Adams2007}
and also \cite[Sec. 2.3]{Basile:2020gqi})
is the equality between two multiplicities:
for any $[m]\otimes [n]$,
\begin{equation}
    N^{m,n}_{\sigma(s,\mu)}=\tilde N^{s,\mu}_{m,n}\,.
\end{equation}
Here, $\sigma(s,\mu)$ is the label
of the $Sp(1,1)$ irrep dual to 
the $O^*(4)$ irrep $\tilde\pi_{s,\mu}$. 

Now let us identify 
the multiplicities $\tilde N^{s,\mu}_{m,n}$. 
The decomposition \eqref{dual dec}
comes in two parts:
the decomposition of the $SU(2)$ irreps,
\begin{equation}
    [m] \otimes [n] = [|m-n|]
    \oplus [|m-n|+2]
    \oplus \cdots \oplus [m+n]\,,
\end{equation}
and the decomposition of the $SL(2,\mathbb R)$
irreps \cite{Repka1978} (see also \cite{Kitaev:2017hnr}),
\begin{equation}
         \cD^-_{m+2}\otimes \cD^+_{n+2}
         = \int_0^\infty {\rm d}\rho\ 
        \cC_{-\frac12+i\,\rho}^{(-1)^{m+n}}\oplus 
        \bigoplus_{0 \leq k < \frac{|m-n|}2}\,
        \cD^{{\rm sgn}(m-n)}_{|m-n|-2k}\,.
    \label{pm decomp}
\end{equation}
We see that the multiplicities are either $1$ or $0$.
Hence, for a fixed $\tilde\pi_{s,\mu}$
the above decomposition simply restricts the possible 
$[m] \otimes [n]$ which appear in the decomposition 
\eqref{dec} of $\pi_{\sigma(s,\mu)}$. Moreover,
we find that certain $\tilde\pi_{s,\mu}$'s do not admit 
any $[m] \otimes [n]$ implying that such irreps
cannot correspond to any (even trivial) $Sp(1,1)$ irrep. 
In other words, they are simply not contained
in the SH representation. Let us see the details now. By choosing the $SU(2)$ irrep as $[2s]$,
$m$ and $n$ are restricted as
\be 
    |m-n|\le 2s \le m+n\,,\qquad 
    m+n-2s\in 2\,\mathbb Z\,.
\ee 
For the $SL(2,\mathbb R)$ irreps with label $\mu$,
we have three choices,
the principal series
$\cC^{\pm}_{\mu=-\frac12+i\,\rho}$,
the complementary series $\cC_\mu$
and the discrete series $\cD^\pm_{2\mu+2}$.
We notice already
that the complementary
series is not available
since it does not appear in the content of the tensor product decomposition,
that is, in the RHS of \eqref{pm decomp}.

If we select a principal series representation 
$\cC_{-\frac12+i\,\rho}^{(-1)^{m+n}}$,
we do not have further restrictions
on possible values of $m$ and $n$. Therefore, we find
\begin{equation}
    \pi_{\sigma(s,-\frac12+i\,\rho)}
    = \bigoplus_{\substack{|m-n| \leq 2s \leq m+n\\
    m+n-2s\in 2 \mathbb Z}} [m] \otimes [n]\,.
\end{equation}
These correspond to 
the spin $s$ principal series representations
of $Sp(1,1)$, describing massive spin $s$ fields.
 
If we select a discrete representation $\cD^{\pm}_{2\mu+2}$,
we find a further restriction on the space and obtain
\begin{equation}
    \pi^\pm_{\sigma(s,\mu)}
    =\bigoplus_{\substack{|m-n| \leq 2s \leq m+n\\
    m+n-2s\in 2 \mathbb Z\\ 2\mu+2 \leq |m-n| \\
    \pm (m-n)>0}} [m] \otimes [n]\,.
\end{equation}
The additional bound on $m$ and $n$ restricts also
possible values of $\mu$.
For integer $s$, we find
$\mu=0,1,\ldots, s-1$,
and for half-integer $s$, we find
$\mu=-\frac12,\frac12,\ldots, s-1$.
These irreps correspond to
the spin $s$ discrete series representation
of $Sp(1,1)$ describing partially-massless spin $s$ fields
and the 
lightest massive fermions.
One can also see that they
always come with two chiralities or helicities $\pm$.

To summarize,
we find that the SH representations
contain exactly all
the unitary representations of $Sp(1,1)$ except for 
the complementary series ones:
the $Sp(1,1)$ (not $SL(2,\mathbb R)$) complementary
series representation correspond to the interval 
$-\frac12\le \mu<1$ for $s=0$ and $-\frac12\le \mu<0$
for $s=1,2,\ldots$ respectively,
while fermions do not appear in the complementary series.
Interestingly, the SH representation
with the dual pair $\big(Sp(1,1),O^*(4)\big)$
contains also the massless spin $s$ fields 
which can be realized
by the $\big(Sp(1,1),O^*(2)\big)$ dual pair.
The conformal scalar with $\mu=0$ (equivalently $\mu=-1$) is in the field content of Vasiliev's higher spin gravity,
together with all integer spin massless fields. 
This conformal scalar in dS$_4$
can be realized only by the latter dual pair.
For more formal treatment
of the $\big(Sp(1,1), O^*(4)\big)$ dual pair,
one may consult with \cite{Li2003, Bao2015}.

\subsection{AdS$_4$}

The $\Lambda<0$ case is more straightforward,
and it is recently discussed in \cite{Basile:2020gqi}.
We use the seesaw diagram,
\begin{equation}
    \parbox{150pt}{
    \begin{tikzpicture}
    \draw [<->] (0,0) -- (1,0.8);
    \draw [<->] (0,0.8) -- (1,0);
    \node at (-1.5,1.2) {$Sp(4,\mathbb R)$};
    \node at (-1.5,0.4) {$\cup$};
    \node  at (-1.5,-0.4) {$U(2)$};
    \node at (2.5,-0.4) {$O(2N)$};
    \node at (2.5,0.4) {$\cup$};
    \node at (2.5,1.2) {$U(2N)$};
    \end{tikzpicture}}
\end{equation}
relating the  reductive dual pairs
$\big(Sp(4,\mathbb R), O(2N)\big)$
and $\big(U(2), U(2N)\big)$ in $Sp(8N,\mathbb R)$.

For $N=1$, the irreps of $O(2)$ are $[2s]_{O(2)}$ 
with $2s\in \mathbb N$ and $[1,1]_{O(2)}$\,.
The one-dimensional irreps $[0]_{O(2)}$
and $[1,1]_{O(2)}$ corresponds
to the scalar irreps of $Sp(4,\mathbb R)$,
whereas $[2s]_{O(2)}$ correspond to the massless
spin $s$ irreps of $Sp(4,\mathbb R)$.
The latter irreps are two dimensional,
composed of the helicity $\pm s$ irreps,
which are related by the $\mathbb Z_2$ part
of $O(2) \cong \mathbb Z_2 \ltimes SO(2)$,
so they assemble into a single irrep for $O(2)$.

For $N=2$, the dual representation of
$[\mu+s,\mu-s]_{O(4)} = [s]_{O(3)} \otimes [\mu]_{O(3)}$
is the discrete series representation
$\cD_{Sp(4,\mathbb R)}(\mu+2,s)$ with 
the lowest energy $\mu+2 = s+2, s+3, \ldots$.
Note that in this case
the SH representation
contains all the massive fields
while excludes the massless fields,
which can be realised
by the $\big(Sp(4,\mathbb R), O(2)\big)$ dual pair. 

Above, we had mentioned that $Sp(4,\mathbb R)$ contains
many representations other than the  discrete series ones.
These irreps would correspond to rather exotic fields such as
tachyon, continuous spin \cite{Metsaev:2016lhs, Metsaev:2017ytk,Metsaev:2019opn}
and even the ones living in 
bitemporal counterpart of AdS$_4$
(see \cite{Basile:2023vyg} for related discussions).
These irreps might be also realized using proper SH representations, namely dual pairs with
different dual groups
$O(1,1)$, $O(2,1)$,
$O(3,1)$ and $O(2,2)$.
In the simplest $O(1)$ case,
the dual pair describes
the conformal scalar and spinor fields in $3d$.
Let us remark also that
this different signature variety is not available for dS$_4$ with $Sp(1,1)$
since $O^*(2N)$ does not allow
any signature variations
and $2N$ must be even.

\subsection{Conformal group}
\label{sec:singletons}
As we had commented above, the four-dimensional
conformal group $\so(2,4)$
\cite{Angelopoulos:1980wg, Angelopoulos:1997ij, 
Laoues:1998ik, Angelopoulos:1999bz}
has a special representation called `singleton'
which reduces to the massless irreps of (A)dS$_4$
with multiplicity one.\footnote{In fact,
the scalar  irrep of $\so(2,4)$ reduces
into two irreps of $\so(2,3)$
which can be interpreted as 
the two possible boundary conditions
of the AdS$_4$ scalar field.}
This can be easily seen from the dual pair
correspondence
\cite[Sec. 8.2]{Basile:2020gqi}.
First, within the SH representation,
the conformal symmetry $SU(2,2)$
that the massless fields enjoy 
is enhanced to $U(2,2)$
with the dual group $U(1)$.
The dS$_4$ group reduction can be 
understood from the dual pairs,
\begin{equation}
    \parbox{150pt}{
    \begin{tikzpicture}
    \draw [<->] (0,0) -- (1,0.8);
    \draw [<->] (0,0.8) -- (1,0);
    \node at (-1.5,1.2) {$U(2,2)$};
    \node at (-1.5,0.4) {$\cup$};
    \node  at (-1.5,-0.4) {$Sp(1,1)$};
    \node at (2.5,-0.4) {$U(1)$};
    \node at (2.5,0.4) {$\cup$};
    \node at (2.5,1.2) {$O^*(2)$};
    \end{tikzpicture}}
\end{equation}
where the reduction of $O^*(2)\cong U(1)$
to $U(1)$ is trivial, thereby
explaining the singleton property
of the massless $Sp(1,1)$ irrep.
Similarly, the AdS$_4$ group reduction
follows the dual pairs,
\begin{equation}
    \parbox{150pt}{
    \begin{tikzpicture}
    \draw [<->] (0,0) -- (1,0.8);
    \draw [<->] (0,0.8) -- (1,0);
    \node at (-1.5,1.2) {$U(2,2)$};
    \node at (-1.5,0.4) {$\cup$};
    \node  at (-1.5,-0.4) {$Sp(4,\mathbb R)$};
    \node at (2.5,-0.4) {$U(1)$};
    \node at (2.5,0.4) {$\cup$};
    \node at (2.5,1.2) {$O(2)$};
    \end{tikzpicture}}
\end{equation}
where again $O(2)\cong U(1)\rtimes \mathbb Z_2$
reduces to $U(1)$ trivially 
except for the scalar case,
and hence the same mechanism
works for the massless $Sp(4,\mathbb R)$ irreps.

\subsection{Other dimensions}
The SH formalism for massless fields in Mink$_4$
can be extended to  3d \cite{Agarwal:2008pu}, 
5d \cite{Chiodaroli:2022ssi},
6d \cite{Cheung:2009dc}
and 
10d \cite{Caron-Huot:2010nes}. 
In case of 3 and 6 dimensions, such SH representations
can be uplifted to the irreps of conformal groups
$\widetilde{SO}^\uparrow(2,3) \cong Sp(4,\mathbb R)$
and $\widetilde{SO}^\uparrow(2,6) \cong O^*(8)$.
Together with the four-dimensional one
$\widetilde{SO}^\uparrow(2,4) \cong SU(2,2)$,
the conformal groups can be regarded as
symplectic groups 
$\mathsf{Sp}(4,\mathbb F)$\footnote{Here,
the symplectic group $\mathsf{Sp}(4,\mathbb F)$
is defined as the matrices $A \in GL(4,\mathbb F)$
satisfying $A^\dagger\,\O_{\sst (4)}\,A=\O_{\sst (4)}$
where $\dagger$ is the conjugation with respect to $\mathbb F$ 
and $\O_{\sst(4)}$ is the four-dimensional
symplectic matrix \cite{Howe:1992bv}.
This definition differs from the standard definition
of symplectic groups.}
over $\mathbb F=\mathbb R, \mathbb C$ and $\mathbb H$,
\begin{equation}
    \mathsf{Sp}(4,\mathbb R)=Sp(4,\mathbb R)\,,
    \qquad 
    \mathsf{Sp}(4,\mathbb C)\cong U(2,2)\,,
    \qquad 
    \mathsf{Sp}(4,\mathbb H) \cong O^*(8)\,.
\end{equation}
These groups naturally include as subgroups
the 3, 4 and 6 dimensional Lorentz groups isomorphic to 
$SL(2,\mathbb R)$, $SL(2,\mathbb C)$ and 
$SL(2, \mathbb H)$, respectively.

For the SH representations of (A)dS fields, 
the (A)dS groups in the spinor representation
need to contain the Lorentz group in the spinor representation.
In 4 dimensions, this was possible thanks to the embedding of the Lorentz group $Sp(2,\mathbb C)$ into $Sp(4,\mathbb R)$ as well as $Sp(1,1)$. We can summarize the situation by the following diagram where
the middle column corresponds to the Lorentz group and its dual, while the left and right columns
correspond to the AdS$_4$ and dS$_4$ groups and their duals, respectively.
\begin{equation}
    \parbox{200pt}{
    \begin{tikzpicture}
    \draw [<->] (-3,0) -- (-3,0.5);
    \draw [<->] (0,0) -- (0,0.5);
    \draw [<->] (3,0) -- (3,0.5);
    \node at (-3,1) {$Sp(4,\mathbb R)$};
    \node at (1.5,1) {$\subset $};
    \node  at (0,1) {$Sp(2,\mathbb C)$};
    \node at (-1.5,1) {$\supset $};
    \node at (3,1) {$Sp(1,1)$};
    \node at (-3,-.5) {$O(2n)$};
    \node at (1.5,-.5) {$\supset $}; 
    \node  at (0,-.5) {$O(2n,\mathbb C)$};
    \node at (-1.5,-.5) {$\subset $};
    \node at (3,-.5) {$O^*(2n)$};
    \end{tikzpicture}}
\end{equation}
In 3 dimensions, we find an analogous structure
which ensures the SH representations of (A)dS$_3$ fields.
The relevant diagram is the following.
\begin{equation}
    \parbox[c]{250pt}{
    \begin{tikzpicture}
    \draw [<->] (-4,0) -- (-4,0.5);
    \draw [<->] (0,0) -- (0,0.5);
    \draw [<->] (3,0) -- (3,0.5);
    \node at (-4,1) {$Sp(2,\mathbb R) \times Sp(2,\mathbb R)$};
    \node at (1.5,1) {$\subset $};
    \node  at (0,1) {$Sp(2,\mathbb R)$};
    \node at (-1.5,1) {$\supset $};
    \node at (3,1) {$Sp(2,\mathbb C)$};
    \node at (-4,-.5) {$O(n) \times O(n)$};
    \node at (1.5,-.5) {$\supset $}; 
    \node  at (0,-.5) {$O(2n)$};
    \node at (-1.5,-.5) {$\subset $};
    \node at (3,-.5) {$O(n,\mathbb C)$};
    \end{tikzpicture}}
\end{equation}
In five dimensions,  we find the following structure
($SU^*(4)\cong \widetilde{SO}^\uparrow(1,5)$ is
the dS$_5$ group).
\begin{equation}
    \parbox{200pt}{
    \begin{tikzpicture}
    \draw [<->] (-3,0) -- (-3,0.5);
    \draw [<->] (0,0) -- (0,0.5);
    \draw [<->] (3,0) -- (3,0.5);
    \node at (-3,1) {$U(2,2)$};
    \node at (1.5,1) {$\subset $};
    \node  at (0,1) {$Sp(1,1)$};
    \node at (-1.5,1) {$\supset $};
    \node at (3,1) {$U^*(4)$};
     \node at (-3,-.5) {$U(2n)$};
  \node at (1.5,-.5) {$\supset $}; 
    \node  at (0,-.5) {$O^*(4n)$};
    \node at (-1.5,-.5) {$\subset $};
    \node at (3,-.5) {$U^*(2n)$};
    \end{tikzpicture}}
\end{equation}
Note that the flat limit of the above 
should agree with the 5d SH representations
constructed in \cite{Chiodaroli:2022ssi}.

\section{Multilinear invariants}
\subsection{Generalities}
The (A)dS$_4$ SH representation 
can be utilized in physical observables 
like scattering amplitudes in flat space. Of course,  
$n$-particle scattering amplitudes in (A)dS$_4$
would not make a literal sense,
and one should regard them rather
as boundary $n$-point correlation functions.
See e.g.  \cite{Baumann:2020dch, Caron-Huot:2021kjy, Jain:2021vrv, Jain:2022ujj, Skvortsov:2022wzo}
for the recent application of SH formalism 
to CFT correlators.
At the technical level, they are nothing but 
the functions of $n$ helicity spinors
invariant under $\mathfrak{Sym}_{\L}$,
which is essentially the branching rule
under the restriction
$\mathfrak{Sym}_{\L}^{\times n}
\downarrow \mathfrak{Sym}_{\L}$\,.
This leads to the dual pair
\begin{equation}
\parbox{300pt}{
\begin{tikzpicture}
\draw [<->] (0,0) -- (1,0.8);
\draw [<->] (0,0.8) -- (1,0);
\node at (-2.5,1.2) {$\mathfrak{Sym}_{\L}\times 
\cdots \times \mathfrak{Sym}_{\L}$};
\node at (-2.5,0.4) {$\cup$};
\node  at (-2.5,-0.4) {$\mathfrak{Sym}_{\L}$};
\node at (3.5,-0.4) {$\mathfrak{Dual}^{(N_1)}_{\L}\times \cdots\times \mathfrak{Dual}^{(N_n)}_{\L}$};
\node at (3.5,0.4) {$\cup$};
\node at (3.5,1.2) {$\mathfrak{Dual}^{(N_1,\ldots,N_n)}_{\L}$};
\end{tikzpicture}}
\label{scattering seesaw}
\end{equation}
where $\mathfrak{Dual}^{(N_1,\ldots,N_n)}_{\L}$ is given by 
\ba
    \mathfrak{Dual}^{(N_1,\ldots,N_n)}_{\L>0}\eq O^*(2(N_1+\cdots+N_n))\,,
    \nn \mathfrak{Dual}^{(N_1,\ldots,N_n)}_{\L<0}\eq O(2(N_1+\cdots+N_p),2(N_{p+1}+\cdots+N_n))\,,
\ea
where $p$ and $n-p$ are
respectively the number of incoming
and outgoing particles. 
Note that
in dS$_4$ case, there is no distinction between incoming and outgoing particles as the energy of a particle is not a conserved quantity.

In \eqref{scattering seesaw}, we require
that the down-right factor
$\mathfrak{Dual}^{(N_1)}_{\Lambda}
    \times \cdots \times 
    \mathfrak{Dual}^{{(N_n)}}_{\Lambda}$
carry an irrep correspondingly to the particle species entering the scattering,
and the down-left $\mathfrak{Sym}_\Lambda$ carry
the trivial representation,
that is invariance under (A)dS$_4$ symmetry.
The translation invariance
condition is deformed
by the derivative part
in $P_{a \dot b}$ \eqref{Poincare},
and becomes more involved to solve,
while the Lorentz invariance
can be easily achieved,
like in the flat space case,
by assuming that the amplitude
is a function of 
the contracted variables,
\begin{equation}
    \langle iI\,jJ\rangle 
    = \lambda^{iI}{}_a\,\lambda^{jJ\,a}\,,
    \qquad 
    [iI\,jJ]
    = \tilde \lambda_{iI\, \dot a}\,
    \tilde\lambda_{jJ}{}^{\dot a}\,.
\end{equation}
Here, $iI, jJ$ are collective indices in which
 $i,j=1,2, \ldots, n$ 
label the particles
entering to the scattering,
whereas $I=1,2,\ldots, N_i$
and $J=1,2,\ldots, N_j$
are the dual group indices
for each particle.
The $\mathfrak{Dual}^{(N)}_{\Lambda}$ irrep 
condition depends on $N$, and it is sufficient
for us to consider $N=1$ and $N=2$.
For $N=1$, it is the usual helicity condition.
For $N=2$ with $\mathfrak{dual}^{{(2)}}_\Lambda
= \mathfrak{s} \oplus \mathfrak{m}_\Lambda$,
the irrep condition of $\mathfrak{s}$
can be imposed like in the flat space case
as in \cite{Conde:2016vxs, Arkani-Hamed:2017jhn},
and we need to impose the irrep condition
of $\mathfrak{m}_\Lambda$ which becomes involved
due to the derivative parts of $M$ and $\tilde M$
given in \eqref{dual alg}.

As a side remark, let us point out
that the complex positive Grassmannian structure
of scattering amplitudes of $n$ massless fields  \cite{Arkani-Hamed:2012zlh, Arkani-Hamed:2013jha, Arkani-Hamed:2013kca}
naturally appears within the framework
of the dual pair correspondence,
as explained in \cite[Sec. 7]{Basile:2020gqi}.
When the scattering  particles
are all massless, that is $N_1=\cdots=N_n=1$, 
the spacetime symmetry $\mathfrak{Sym}_\L$ 
is enhanced to $U(2,2)$, while the dual group 
$\mathfrak{Dual}^{(1,\ldots,1)}_\Lambda$
becomes the indefinite unitary group $U(p,n-p)$
in the dual pairs \eqref{scattering seesaw}.
In this enhanced setting,
we do not require the full invariance under $U(2,2)$
but only
under the subgroup $\mathfrak{Sym}_\Lambda$,
which contains the Lorentz subgroup $SL(2,\mathbb C)$.
Together with the diagonal subgroup $\mathbb C^\times$
generated by the total helicity
and the dilation operator,
the Lorentz $SL(2,\mathbb C)$ can be uplifted 
to $GL(2,\mathbb C)$,
which has $GL(n,\mathbb C)$ as its dual group.
The situation can be 
again summarized by
the following seesaw diagram.
\begin{equation}
    \parbox{250pt}{
    \begin{tikzpicture}
    \draw [<->] (0,-0.8) -- (1.5,0.8);
    \draw [<->] (0,0) -- (1.5,0);
    \draw [<->] (0,0.8) -- (1.5,-0.8);
    \node at (-2,1.4) {$U(2,2) \times \cdots \times U(2,2)$};
    \node at (-2,0.7) {$\cup$};
    \node  at (-2,0) {$U(2,2)$};
    \node at (-2,-0.7) {$\cup$};
    \node at (-2,-1.4) {$GL(2,\mathbb C)$};
    \node at (3.5,1.4) {$GL(n,\mathbb C)$};
    \node at (3.5,0.7) {$\cup$};
    \node  at (3.5,0) {$U(p,n-p)$};
    \node at (3.5,-0.7) {$\cup$};
    \node at (3.5,-1.4) {$U(1) \times \cdots \times U(1)$};
    \end{tikzpicture}}
\end{equation}
The Lorentz invariance is equivalent to
the condition that 
under restriction to $GL(2,\mathbb C)$,
the amplitudes carry a one-dimensional
representation, wherein $SL(2,\mathbb C)$
acts trivially,
and $GL(1,\mathbb C) \cong \mathbb C^\times$
acts diagonally.
The corresponding $GL(n,\mathbb C)$ 
representation is a degenerate 
principal series representation
(see e.g. \cite{Howe1999}),
which is realized as the space of functions
on the complex positive Grassmannian manifold $Gr_{2,n}(\mathbb C)$.

Coming back to the picture
\eqref{scattering seesaw}, 
the only non-trivial
part of the conditions are the translational
invariance condition, and the irrep condition
of $\mathfrak{m}_\Lambda$ for $N=2$. 
When $\Lambda=0$, both of these conditions
are algebraic and could be solved by imposing
the helicity spinors to be constrained
on the shell of the momentum conservation
and constant mass-squared.
When $\Lambda\neq0$, both of these conditions
become differential equations.

\subsection{Translational invariance}
Let us consider first the condition
of translation invariance,
\begin{equation}
    P_{a \dot b}\,\cA = \left(\lambda^{\cI}{}_{a}\,
    \tilde \lambda_{\cI\,\dot b}
    +\Lambda\,\frac{\partial}{\partial \lambda^{\cI\,a}}
    \frac{\partial}{\partial \tilde{\lambda}_{\cI}{}^{\dot b}}\right)\cA=0\,,
    \label{p inv}
\end{equation}
where $\cI=iI,\cJ=jJ$ 
denote the collective indices.
In the Mink$_4$ case, the solution is 
nothing but the momentum conservation delta distribution $\delta^4(p)$ with
\begin{equation}
    p_{a \dot b} =\lambda^{\cI}{}_{a}\,
    \tilde \lambda_{\cI\,\dot b}\,,
\end{equation}
and hence we expect a similar kind of 
distributional property 
for the $\Lambda\neq0$ solution.
For the massless 3pt case, 
this equation has been
analyzed in detail in
\cite[App. E]{Nagaraj:2019zmk},
where the authors made an ansatz
as a function of $\langle 12 \rangle [12]$,
$\langle 23 \rangle [23]$ and $\langle 31 \rangle [31]$
and derived a system of four PDEs.
Instead of solving these equations directly,
they checked that the amplitudes obtained
from field theoretical approach (that is,
spacetime integral of three AdS plane wave solutions)
solve the equations. The solution is spanned
by four independent distributions of 
$\langle 12 \rangle\,[12] +\langle 23 \rangle\,[23]
+ \langle 31 \rangle\,[31]
= \tfrac12\,p_{a\dot b}\,p^{a\dot b}$.

Let us revisit the problem slightly differently
for the general case (massive or massless $n$-pt).
Since $\cA$ should involve
the momentum conservation delta function in the flat limit, we consider the ansatz $\cA=\cA(p_{a\dot b},
    \langle \cI\cJ\rangle , [\cI\cJ])$\,.\footnote{Note
that the variables $p_{a\db}$
are not independent from 
$\la \cI\cJ\ra$ and $[\cI\cJ]$,
as they are related by
$p_{a\db}\,p^{a\db}
=\la \cI\cJ\ra [\cI\cJ]$.
Therefore, 
whenever the latter combination appears, we have to regard them as a function of $p_{a\dot b}$
to avoid the related ambiguities.
}
Then the condition \eqref{p inv} sets up the differential equation,
\begin{equation}
    \left[p_{a\dot b} + \Lambda\left(p^{c\dot d}\,
    \frac{\partial}{\partial p^{{a\dot d}}}\,
    \frac{\partial}{\partial p^{{c\dot b}}}
    + H\,\frac{\partial}{\partial p^{a\dot b}}
    + \lambda^{\cJ}{}_a\,\tilde\lambda_{\cK\dot b}
    \frac{\partial}{\partial \langle \cI\cJ\rangle}
    \frac{\partial}{\partial [\cI\cK]}\right)\right]\cA=0\,,
    \label{p diff}
\end{equation}
with  the number operator $H$,
\begin{equation}
    H=\cN+\frac12\,\langle\cI\cJ\rangle\,
    \frac{\partial}{\partial\langle\cI\cJ\rangle}
     +\frac12\,[\cI\cJ]\,
     \frac{\partial}{\partial[\cI\cJ]}\,,
\end{equation}
where $\cN=\sum_{i=1}^n N_i$ is the sum
of the ranks of the dual groups
for all $n$ particles,
and the factor $1/2$ has been introduced 
to take the antisymmetry of $\cI\cJ$ into account.
The last term of the differential equation \eqref{p diff}
is problematic since it is \emph{not} expressed in terms
of the variables $p_{a \dot b}$, $\langle\cI\cJ\rangle$
and $[\cI\cJ]$ only.
We can bypass the problem by 
focusing on the ``longitudinal part''
of the equation:
contracting \eqref{p diff} with $p^{a\db}$,
we find 
\begin{equation}
    \left[p^{a \dot b}\,p_{a \dot b}
    + \Lambda\left(p^{a \dot b}\,p^{c\dot d}\,
    \frac{\partial}{\partial p^{a \dot d}}\,
    \frac{\partial}{\partial p^{c \dot b}}
    + H\,p^{a \dot b}\,
    \frac{\partial}{\partial p^{a \dot b}}
    + 2\,R\right)\right]\cA=0\,,
    \label{p diff eq}
\end{equation}
where $R$ is  a differential operator acting
on the Lorentz invariant variables as
\be
   R = \frac12\,\la\cI\cJ\ra\,
   \frac{\partial}{\partial \langle \cJ\cL\rangle}\,
   [\cI\cK]\,\frac{\partial}{\partial [\cK\cL]}\,.
\ee 
Viewing $\Lambda$ as a deformation parameter, 
our aim is to find the deformation
of the delta distribution solution
of the $\Lambda=0$ case.
We can better control the situation by
going to the Fourier space
$q^{a \dot b}$ where
the constant solution corresponds to the correct delta distribution in the $\Lambda=0$ case.
Since the constant solution is isotropic,
we assume that 
$\tilde\cA$ is a function of 
$t=\frac12\,q^{a \dot b}\,q_{a \dot b}$,
and this reduces the equation
to the simple second order differential equation
in the $t$ variable,
\begin{equation}
    \left[\left(t\,\frac{\partial}{\partial t} + 2\right)\,
    \left((1-\Lambda\,t)\,\frac{\partial}{\partial t}
    +\Lambda\,(H-4)\right) -\Lambda\,R\right]\,
    \tilde\cA = 0\,.
    \label{2nd t diff}
\end{equation}
We can use the separation of variables, 
\begin{equation}
    R\,\tilde\cA_r = r\,\tilde\cA_r\,,
\end{equation}
to decompose the PDE \eqref{2nd t diff}
into hypergeometric differential equations
with two types of solutions, the first one being,
\begin{equation}
    \tilde \cA_r ={}_2F_1\Big(a_+,a_-,2\,;\,\Lambda\,t\Big)\,f_r\,,
    \label{trans sol}
\end{equation}
where $a_\pm$ are
\be
    a_\pm=\frac12\left(6-H\pm\sqrt{(H-2)^2-r}\right),
\ee 
and $f_r$ is an arbitrary function
of $\la \cI\cJ\ra$ and $[\cI\cJ]$.
The second solution of the hypergeometric
differential equation takes the form,
\be 
    \frac1{t}
    +\L\,\sum_{n=0}^\infty
    \frac{(a_+-1)_{n+1}\,(a_--1)_{n+1}}{(n+1)!\,n!}\left[\ln (\L\,t)\,
    (\L\,t)^n+c_n\,(\L\,t)^{n+1}\right]\,,
    \label{2nd sol}
\ee
with 
\be
    c_n=
    \sum_{m=0}^{n-1}
    \left(
    \frac1{a_++m}
    +\frac1{a_-+m}
    -\frac1{m+2}
    -\frac1{m+1}\right).
\ee
The hypergeometric function ${}_2F_1(a_+,a_-,2;\L\,t)$ reduces to $1$
for $\L=r=0$,
while the second solution \eqref{2nd sol} to $1/t$.
Since the constant solution corresponds to the desired delta distribution,
we retain only the 
hypergeometric function.
Remark that for $r=0$, 
the hypergeometric function gets simplified to give
\be 
    \tilde \cA_0
    =(1-\L\,t)^{H-4}\,f_0\,.
\ee
This  is consistent with the expressions
obtained in \cite{Nagaraj:2018nxq, Nagaraj:2019zmk, 
Nagaraj:2020sji} for massless 3pt.
Remark also that
the hypergeometric function
\eqref{trans sol}
 has a branch point at $\L\,t=1$,\footnote{In \cite{Nagaraj:2018nxq, Nagaraj:2019zmk, 
Nagaraj:2020sji}, 
the $q_{a\db}$ variables
carry a spacetime coordinate
interpretation, and
the branch point corresponds to the boundary of the coordinate chart.}
which might be interpreted
as the cosmological horizon
and related to the alpha vacua.\footnote{In de Sitter space,
there is a one-parameter family of dS invariant
vacuum states \cite{Allen:1985ux}. This vacuum
ambiguity would lead to an analogous ambiguity
in $n$-point correlation functions.}
Eventually, the most general invariant will be 
linear combinations of  $\cA_r$
with different $r$ values.

\subsection{Mass condition}
Let us move on to the irrep condition
of $\mathfrak{m}_\Lambda$ for each
of the $n$ particles, fixing their masses.
For the discrete series irreps
of $\mathfrak{dual}^{(2)}_{\Lambda>0}$
in the dS$_4$ case and the finite-dimensional
irreps of $\mathfrak{dual}^{(2)}_{\Lambda<0}$
in the AdS$_4$ case, we can impose
the highest weight condition $M_i\,\cA=0$
or the lowest weight condition $\tilde M_i\,\cA=0$\,,
on the $K_i$ eigenstate with
\begin{equation}
    K_i\,f=\mp2\,(\mu_i+1)\,f\,,
    \label{dS K}
\end{equation}
for dS$_4$ and 
\begin{equation}
    K_i\,f=\pm2\mu_i\,f\,,
    \label{AdS K}
\end{equation}
for AdS$_4$. Here, the $K_i$, upon acting on $f$,
reduces to the differential operator,
\begin{equation}
    K_i\,f=\left(\la iI\,\cJ\ra\,\frac{\partial}{\partial \la iI\,\cJ\ra}
    -[ iI\,\cJ] \,\frac{\partial}{\partial [ iI\,\cJ]}\right)f\,,
\end{equation}
where the repeated indices $\cJ$ and $I$ are summed over except for the particle label $i$.
The highest weight condition $M\,\cA=0$ can be translated as well into the differential equations,
\begin{equation}
    M_i\,f = \left[\langle i1\, i2 \rangle
    +\,\Lambda
    \left(2\,\frac{\partial}{\partial [i1\,i2]}
    + [\cJ\cK]\,\frac{\partial}{\partial [i1\,\cJ]}\,
    \frac{\partial}{\partial [i2\,\cK]}\right)\right]f=0\,,
    \label{M=0}\nonumber
\end{equation}
where the repeated indices $\cJ, \cK$ include $i$'th particle's values $iI$,
and the lowest weight condition 
is simply given by the complex conjugate of the above.

Note that the $K$ eigenstate conditions
\eqref{dS K} and \eqref{AdS K} become singular
in the flat limit where $\mu$ is sent to infinity
while $\mu\sqrt{|\Lambda|}$ held fixed.
Moreover, the principal series irreps
of dS$_4$ have neither a highest 
nor a lowest weight state. 
Therefore, the above conditions are 
inapplicable in that case.
We may consider to use
the $K$ eigenstate with eigenvalue $0$
or $\pm1$ to avoid this problem,
but in that case we cannot use any more
the simple condition $M=0$ (or $\tilde M=0$).
Instead we need to use the Casimir condition involving 
the anticommutator $\{M,\tilde M\}$ resulting
in a fourth-order differential equation instead
of \eqref{M=0}.

In fact, 
for the principal series irreps, 
it is more natural to impose
\begin{subequations}
\begin{eqnarray}
    && (M_i-\tilde M_i)\,f
    =2\,\sqrt{\Lambda}\,\mu_i\,f\,,
    \label{M-M cond}\\
    && (M_i+\tilde M_i-\sqrt{\Lambda}\,K_i)\,f=0\,,
    \label{M+M-K cond}
\end{eqnarray}
\end{subequations}
which has also a well-defined flat limit, 
and can be expressed
as second order differential equations
in $\la \cI\cJ\ra$ and $[\cI\cJ]$.
Solving these conditions is
beyond the scope of the current work.
Instead, let us make a few remarks
on the change of basis where
the $O^*(4)$ actions
become more natural.

For the change of basis,
we fix the convention as $a,b,\da,\db=+,-$
and $\epsilon_{-+}=\epsilon^{+-}=1$
and perform Fourier transform with respect to $\lambda^I_-$ and its complex conjugate as
\be
    \left(\,\frac{\lambda^{I}{}_-}{\sqrt\L}\,,\,
    \sqrt\L\,\frac{\partial}{\partial\lambda^{I}{}_-}\,,\,
    \frac{\tilde\lambda_{I-}}{\sqrt\L}\,,\,
    \sqrt\L\,\frac{\partial}{\partial\tilde\lambda_{I-}}\,\right)
    \ \longrightarrow\  i\left(\,\frac{\partial}
    {\partial\zeta_I}\,,\,
    \zeta_I\,,\,
    \frac{\partial}{\partial\tilde\zeta^I}\,,\,
    \tilde\zeta^I\,\right).
    \label{Fourier}
\ee
Then, the dual algebra generators read
\begin{subequations}  
\begin{eqnarray}
    &&K^I{}_J=
    \lambda^I\,\frac{\partial}{\partial \lambda^J}-
     \zeta_J\,\frac{\partial}{\partial \zeta_I}-
      \tilde\lambda_J\,\frac{\partial}{\partial \tilde \lambda_I}+
       \tilde\zeta^I\,\frac{\partial}{\partial \tilde \zeta^J}\,,
       \label{K lin}
    \\
    &&\frac{M^{IJ}}{\sqrt\L}=i\left(\lambda^I\,\frac{\partial}{\partial \zeta_J}
    -\lambda^J\,\frac{\partial}{\partial \zeta_I}
    +\tilde \zeta^I\,\frac{\partial}{\partial\tilde \lambda_J}
    -\tilde \zeta^J\,\frac{\partial}{\partial \tilde \lambda_I}\right),
    \label{M lin}
\end{eqnarray}
\end{subequations} 
where we used $\lambda^I=\lambda^I_+$
and $\tilde\lambda_I=\tilde\lambda_{I+}$\,.
In this basis,
the $\mathfrak{o}^*(2N)$
generators become
first order differential operators, and hence
can be easily integrated to
a Lie group.
This basis admits in fact
a natural realization
in terms of quaternions: see Appendix \ref{app:quaternion} for the details.
While the new basis
\eqref{K lin} and \eqref{M lin} renders
the dual algebra as simple first-order differential operators,
the Lorentz algebra becomes second-order instead.
In other words, 
in the basis where the dual algebra is linearly realized, the dS$_4$ algebra is not. And vice-versa: we can go to another basis where the dS$_4$ algebra is realized linearly, but then the dual algebra is not. 

For $N=2$, we can consider
a different Fourier transformation,
\be
    \left(\,\frac{\lambda^{2}{}_a}{\sqrt\L}\,,\,
    \sqrt\L\,\frac{\partial}{\partial\lambda^{2}{}_a}\,,\,
    \frac{\tilde\lambda_{2\da}}{\sqrt\L}\,,\,
    \sqrt\L\,\frac{\partial}{\partial\tilde\lambda_{2\da}}\,\right)
    \ \longrightarrow\  i\left(\,\frac{\partial}
    {\partial\xi^a}\,,\,
    \xi^a\,,\,
    \frac{\partial}{\partial\tilde\xi^{\da}}\,,\,
    \tilde\xi^{\da}\,\right),
    \label{Fourier2}
\ee
where only the $I=2$ variables
are transformed.
Upon a further change
of basis,
\be 
    z^a=\frac{\xi^a-i\,\lambda^{1a}}2\,,
    \qquad w^a=\frac{\xi^a+i\,\lambda^{1a}}2\,,
    \qquad 
    \tilde z^{\da}=\frac{\tilde\xi^{\da}+i\,\tilde\lambda_1{}^{\da}}2\,,
    \qquad \tilde w^{\da}=\frac{\tilde\xi^{\da}-i\,\tilde\lambda_1{}^{\da}}2\,,
\ee 
the conditions \eqref{M-M cond} and \eqref{M+M-K cond} 
become simple: 
\begin{subequations}
\begin{eqnarray}
    &&\frac{M-\tilde M}{\sqrt\L}
    =z^a\,\frac{\partial}{\partial z^a}
    +\tilde z^{\da}\,\frac{\partial}{\partial \tilde z^{\da}}
    -w^{a}\,\frac{\partial}{\partial w^a}-\tilde w^{\da}\,\frac{\partial}{\partial \tilde w^{\da}}\,,
    \label{M-M cond'}\\
    && \frac{M+\tilde M}{\sqrt\L}-K
=2\left(z^a\,\frac{\partial}{\partial w^a}
 -\tilde z^{\da}\,\frac{\partial}{\partial \tilde w^{\da}}
 \right).
    \label{M+M-K cond'}
\end{eqnarray}
\end{subequations} 
The condition \eqref{M+M-K cond} can be solved
by an arbitrary function of $z^a$, $\tilde z^{\da}$,
and $z^a\,\tilde w^{\db}+w^a\,\tilde z^{\db}$. Furthermore,
the  condition \eqref{M-M cond}, which fixes
the principal series label,
becomes a simple homogeneity condition with respect to the number operator \eqref{M-M cond'}.
The variables $z^a\,\tilde w^{\db}+w^a\,\tilde z^{\db}$ have weight zero 
and hence are not constrained,
while the homogeneity of  $|z|=\sqrt{z^a\,\tilde z^{\da}}$ is restricted to $\mu$.
The spin condition further restricts
the variables $z^a\,\tilde w^{\db}+w^a\,\tilde z^{\db}$
and $z^a/|z|$.
In the end, the remaining freedom
corresponds to the massive 
 irrep of $Sp(1,1)$.
However,
in this basis,
the spin part $\mathfrak{s}$
of the dual algebra, that is generated by $K^I{}_J$,
is realized by second-order differentials.

Therefore, the dS$_4$ invariance
condition
and the $O^*(4)$ irrep condition
for each of the particles cannot be
solved within a single basis, but
by employing muliple bases that are related by Fourier transformations. These conditions may be solved for concrete examples of interest. We leave this to future investigations.

\acknowledgments{We are grateful to Eduardo Conde for agreeable communications.
The work of T.B. was supported by the European Union’s Horizon 2020 research and innovation program under the Marie Sk\l{}odowska-Curie grant agreement No 101034383. 
E. J. was supported by the National Research Foundation of Korea (NRF) grant funded by the Korea government (MSIT) (No. 2022R1F1A1074977). 
K. M. was supported by the European Union’s Horizon 2020 Research and Innovation Programme under the Marie Sk\l{}odowska-Curie Grant No. 844265, and STFC Consolidated Grants ST/T000791/1 and ST/X000575/1.}

\appendix
\section{UIRs of dS$_4$ group and Casimirs}
\label{app:dixmier}

Unitary and irreducible representations
of the dS$_4$ isometry group 
were first classified by J. Dixmier
in \cite{Dixmier1961}. In this appendix,
we recall this classification and
the eigenvalues of the quadratic and quartic Casimir operators.
\begin{itemize}
    \item 
    $\pi^{\pm}_{p,q}$\,:\,
    [$p=\frac12,1,\frac32,\ldots$ ; 
    $q=p, p-1, \ldots, 1$ or $\frac12$]
    with 
    \begin{eqnarray}
        && C_2 = -p(p+1) - (q-1)q + 2
        = -p(p+1) -(q+1)(q-2)\,,
        \nn  
        && C_4=-p(p+1)(q-1)q\,,
    \end{eqnarray}
    corresponding to the discrete series.
    \item 
    $\pi_{p,0}$\,:\,[$p=1,2,\ldots$]
    with the quadratic and quartic Casimir
    operators taking the values
    \begin{eqnarray}
        && C_2 = p(p+1)-2\,,
       \nn 
        && C_4 = 0\,.
    \end{eqnarray}
    These UIRs form the discrete series.
    \item 
    $\nu_{p,\s}$\,:\,
    [$p=0$ ; $\s>-2$]
    and 
    [$p=1,2,\ldots $ ; $\s>0$]
    and [$p=\frac12,\frac32,\ldots $ ; $\s>\frac14$]
    with 
    \ba
        && C_2=p(p+1)-\s-2\,,
        \nn 
        && C_4=-p(p+1)\,\s\,,
    \ea  
    corresponding to the principal and complementary series.
\end{itemize}

\section{Quaternion
realization of dS$_4$ group}
\label{app:quaternion}
The dual pair $\big(Sp(M,M), O^*(2N)\big)$
can be naturally realized in terms of quaternions.
The oscillator representation 
is the space of functions on $\mathbb H^{MN}$,
where $O^*(2N)$ 
acts on a function $\Phi$ by right multiplication,
\begin{equation}
    \big\langle\mathsf Q\,\big|\,
    U_{O^*(2N)}(\mathsf A)\,\Phi\big\rangle
    =\big\langle\mathsf Q\,\mathsf A\,\big|
    \Phi\big\rangle\,, 
\end{equation}
where $\mathsf Q$ is an $M \times N$ quaternionic matrix
and $\mathsf A$ is an
$N \times N$ quaternionic matrix
satisfying\footnote{Here, $\mathsf j$
denotes the basis element of quarternions
that can be represented by the Pauli matrix
$i\,\sigma_2$.}
\be 
 \mathsf A^\dagger\,\mathsf j\, 
    \mathsf A=\mathsf j\,,
\ee 
thereby representing an arbitrary element
of $O^*(2N)$.
For $M=1$, each of the quaternionic elements
of $\mathsf Q=(\mathsf q_I)$,
seen as a $2\times 2$ complex matrix,
can be parameterized by two complex numbers as
\be
    \mathsf q_I=
    \begin{pmatrix} 
    \lambda^I+i\,\zeta_I & 
    \zeta_I+i\,\lambda^I
    \\ 
    -\tilde \zeta^I+i\,\tilde\lambda_I
    & \tilde \lambda_I
    -i\,\tilde\zeta^I
    \end{pmatrix}\,.
    \label{q para}
\ee 
Note that
we recover the expressions \eqref{K lin} and \eqref{M lin}
from the above parameterization of $\mathsf q_I$. 

For even $N=2L$, the $Sp(M,M)$
action can also be represented by the left multiplication of 
a quaternionic matrix,
\begin{equation}
    \bigg\langle
   \binom{\mathsf Q_1}
    {\mathsf P_2}\,\bigg|\,
    U_{Sp(M,M)}(\mathsf B)\,\Phi\bigg\rangle
    =\bigg\langle
    \mathsf B^t\, \binom{\mathsf Q_1}
    {\mathsf P_2}\,\bigg|\,
    \Phi\bigg\rangle\,, 
\end{equation}
where $\mathsf B$ is
an element
of $Sp(M,M)$, and hence
a $2M\times 2M$
quaternionic matrix satisfying
\be
   \mathsf B^\dagger \begin{pmatrix}0&I_M\\I_M&0\end{pmatrix} 
    \mathsf B=\begin{pmatrix}0&I_M\\I_M&0\end{pmatrix}.
\ee 
The sub-matrices $\mathsf Q_1$ and $\mathsf P_2$ are $M\times L$ quaternionic matrices,
and $\mathsf P_2$ is the Fourier
conjugate of $\mathsf Q_2$ where
$\mathsf Q_1$ and $\mathsf Q_2$ 
form the $M\times 2L$ matrix 
$\mathsf Q=(\mathsf Q_1\,\mathsf Q_2)$.

\bibliography{biblio}
\bibliographystyle{JHEP}

\end{document}